\documentclass[aps,prl,twocolumn,superscriptaddress,floatfix,showpacs]{revtex4-1}
\usepackage{graphicx}
\usepackage{color}

\newcommand{\kap}{Cu$_3$Zn(OH)$_6$Cl$_2$}
\newcommand{\kapp}{(Cu$_{0.73}$Zn$_{0.27}$)$_3$(Zn$_{0.88}$Cu$_{0.12}$)(OH)$_6$Cl$_2$}
\newcommand{\XE}{\ensuremath{\chi''(E)}}
\newcommand{\SQE}{\ensuremath{S(Q,E)}}
\newcommand{\SMQ}{\ensuremath{S_{\rm inel}^{\rm mag}(Q)}}
\newcommand{\etal}{\textit{et al.}}
\newcommand{\Ang}{\AA$^{-1}$}
\newcommand{\uSR}{$\mu$SR}
\newcommand{\XDC}{\ensuremath{\chi_{\rm DC}}}
\newcommand{\Xloc}{\ensuremath{\chi_{\rm loc}}}

\begin{document}

\title{Kapellasite: a kagome quantum spin liquid with competing interactions}
\author{B. F{\aa}k}
   \affiliation{SPSMS, UMR-E CEA/UJF-Grenoble-1, INAC, F-38054 Grenoble Cedex 9, France}
\author{E.~Kermarrec}
\affiliation{Laboratoire de Physique des Solides, Universit\'e Paris Sud 11, UMR CNRS 8502, F-91405 Orsay, France}
\author{L. Messio}
   \affiliation{Institute of Theoretical Physics, \'Ecole Polytechnique F\'ed\'erale de Lausanne,
CH-1015 Lausanne, Switzerland}
\author{B. Bernu}
   \affiliation{LPTMC, %Laboratoire de Physique Th\'eorique de la Mati\`ere Condens\'ee,
UMR 7600 CNRS, Universit\'e Pierre et Marie Curie, Paris VI, F-75252 Paris Cedex 05, France}
\author{C. Lhuillier}
   \affiliation{LPTMC, %Laboratoire de Physique Th\'eorique de la Mati\`ere Condens\'ee,
UMR 7600 CNRS, Universit\'e Pierre et Marie Curie, Paris VI, F-75252 Paris Cedex 05, France}
\author{F.~Bert}
   \affiliation{Laboratoire de Physique des Solides, Universit\'e Paris Sud 11,
UMR CNRS 8502, F-91405 Orsay, France}
\author{P.~Mendels}
   \affiliation{Laboratoire de Physique des Solides, Universit\'e Paris Sud 11,
UMR CNRS 8502, F-91405 Orsay, France}
   \affiliation{Institut Universitaire de France, 103 bd Saint-Michel,
F-75005 Paris, France}
\author{B.~Koteswararao}
   \affiliation{Laboratoire de Physique des Solides, Universit\'e Paris Sud 11,
UMR CNRS 8502, F-91405 Orsay, France}
\author{F.~Bouquet}
   \affiliation{Laboratoire de Physique des Solides, Universit\'e Paris Sud 11,
UMR CNRS 8502, F-91405 Orsay, France}
\author{J.~Ollivier}
   \affiliation{Institut Laue Langevin, BP156, F-38042 Grenoble, France}
\author{A. D.~Hillier}
   \affiliation{ISIS Facility, STFC, Rutherford Appleton Laboratory, Chilton, Didcot, Oxon OX11 OQX, UK}
\author{A.~Amato}
   \affiliation{Laboratory for Muon Spin Spectroscopy, Paul Scherrer Institut, CH-5232 Villigen PSI, Switzerland}
\author{R. H.~Colman}
   \affiliation{University College London, Department of Chemistry, 20 Gordon Street, London, WC1H 0AJ, UK}
\author{A. S.~Wills}
   \affiliation{University College London, Department of Chemistry,  20 Gordon Street, London, WC1H 0AJ, UK}
\date{\today}

\begin{abstract}
Magnetic susceptibility, NMR, $\mu$SR, and inelastic neutron scattering measurements
show that kapellasite, \kap, a geometrically frustrated spin-1/2 kagom\'e antiferromagnet polymorphic with herbertsmithite,
is a gapless spin liquid showing unusual dynamic short-range correlations of non-coplanar cuboc2 type which persist down to 20 mK.
The Hamiltonian is determined from a fit of a high-temperature series expansion to  bulk susceptibility data and possesses competing exchange interactions.
The magnetic specific heat calculated from these exchange couplings is in good agreement with experiment.
The temperature dependence of the magnetic structure factor and the muon relaxation rate are
calculated in a Schwinger-boson approach and compared to experimental results.
\end{abstract}

\maketitle

The quantum spin liquid (QSL)
is one of the most elusive states in condensed matter physics,
even after two decades of intense studies  \cite{Balents10},
with implications ranging from geometrically or exchange frustrated magnetic systems
to unconventional superconductivity.
A promising model system for the observation of the QSL state
is the $S\!=\!1/2$ kagom\'e antiferromagnet (KAFM),
which is characterized by quantum spins
on a geometrically frustrated lattice of corner-sharing equilateral triangles.
The first suitable model material,
the mineral herbertsmithite, is classified as a gapless spin liquid
\cite{Shores05,Mendels07,Helton07,Olariu08,deVries09,Mendels10,Jeong11},
and still lacks a complete understanding.
Its discovery triggered a burst of experimental and theoretical investigations
in the field of quantum spin liquids in two dimensions
and challenges the most advanced theoretical approaches~\cite{Ran07,Yan11,Lu11,Messio12}.

\begin{figure}[b!]
\includegraphics[width=0.98\columnwidth]{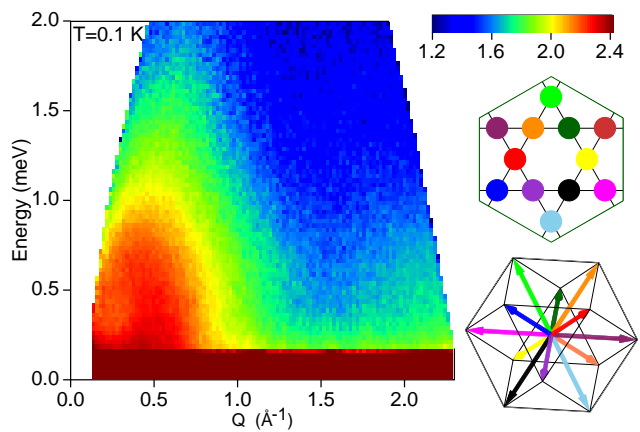}
\caption{(Color online) Neutron scattering intensity, \SQE, on a logarithmic scale
as a function of wave vector $Q$ and energy $E$
of kapellasite at a temperature of 0.1~K
measured with an incoming neutron energy of 3.27 meV.
The diffuse inelastic magnetic scattering is centered at $Q$=0.5 and 2.0 \Ang.
On the right are shown the spin directions (bottom) on the kagom\'e lattice (middle) of the cuboc2 magnetic structure.}
\label{FigMap}
\end{figure}

A new experimental realization of the $S\!=\!1/2$ KAFM model
was recently synthesized \cite{Colman08,Colman10},
the mineral kapellasite, \kap,
a polymorph of herbertsmithite,
i.e., they have the same chemical formula but different crystallographic structures.
The layered structure of kapellasite prevents the
Cu/Zn disorder present in these materials from either
coupling the kagom\'e planes or creating free spins,
in contrast to the more 3-dimensional herbertsmithite.
In this Letter, we use inelastic neutron scattering and muon spin relaxation (\uSR) down to 20~mK
to show that kapellasite is a gapless spin liquid
without any sign of long-range magnetic order or spin freezing.
In contrast to the featureless excitation spectrum of herbertsmithite,
kapellasite exhibits novel dynamic short-range correlations with a well-defined wave-vector  dependence
(see Fig.\ \ref{FigMap}),
reminiscent of the non-coplanar twelve-sublattice cuboc2 magnetic structure
\cite{Domenge05},
where, due to a particular set of ferro- and antiferromagnetic competing interactions,
spins on a given hexagon are coplanar with
neighboring spins forming an angle of 60$^\circ$ with each other;
the third-neighbor spins across hexagons are antiparallel.

% FIG. 1. Experimental Map

\begin{figure}[t!]
\includegraphics[width=0.8\columnwidth]{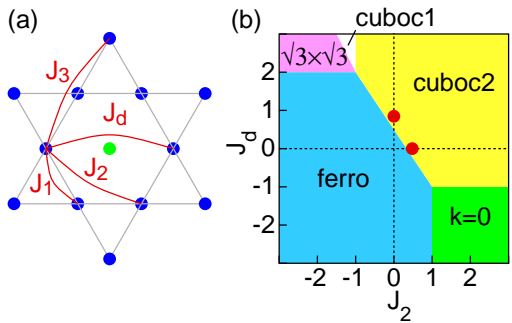}
\caption{(Color online) (a) Kagom\'e plane of kapellasite
with Cu$^{2+}$ $S\!=\!1/2$ spins (blue),
non-magnetic Zn$^{2+}$ ion (green),
and exchange interactions (red).
(b) Classical $J_2$--$J_d$ phase diagram for ferromagnetic nearest-neighbor coupling,
$J_1=-1$, and $J_3=0$.
The best two-parameter HT-series fits to susceptibility data
are shown with red points.}
\label{FigLattice}
\end{figure}

Kapellasite crystallizes in the $P\bar{3}m1$ trigonal space group (\#164) with
lattice parameters $a$=6.33 and $c$=5.70~\AA.
The $S$=1/2 kagom\'e lattice of undistorted corner-sharing triangles
is obtained by regularly doping a two-dimensional (2D) triangular Cu$^{2+}$ metal-site sublattice
with diamagnetic Zn$^{2+}$ ions [see Fig.\ \ref{FigLattice}(a)].
Inductively coupled plasma mass spectrometry
showed that our sample has not the ideal Cu/Zn 3:1 stoichiometry
and $^{35}$Cl NMR showed some Cu/Zn intersite mixing~\cite{Kermarrec12}.
The actual chemical formula consistently determined with neutron powder diffraction \cite{Colman10}
is  \kapp,  with 27\% Zn on the Cu-sites of the kagom\'e lattice
and 12\% Cu on the ``hexagonal'' Zn site.
The Cu/Zn mixing leads to some disorder {\it within} the kagom\'e planes,
but cannot induce any coupling between the planes,
which occurs only via very weak O-H-Cl hydrogen bonds  \cite{Colman08}.
Kapellasite is therefore remarkably two-dimensional.

% FIG. 2. Phase diagram

\begin{figure}[t!]
\center
\includegraphics[scale=0.3]{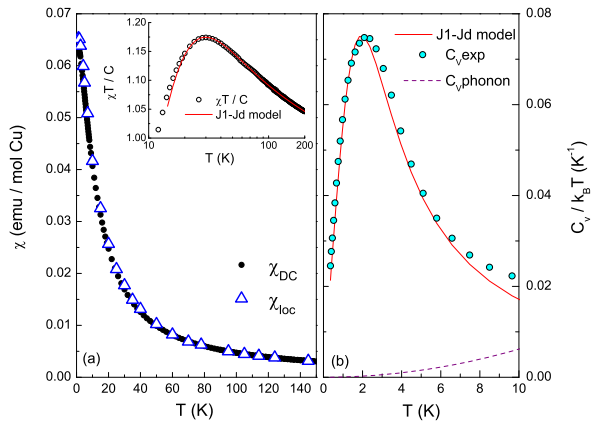}
\caption{(Color online) (a) Local  \Xloc\ and macroscopic \XDC\ magnetic susceptibilities
from NMR and DC SQUID measurements, respectively.
Inset: Fit of high-temperature series expansion (red line) down to 20 K.
(b) Total specific heat per spin $C_{\rm V}^{\rm exp}$ measured in zero field (symbols),
magnetic specific heat calculated with the $J_1$--$J_d$ model
(red solid line)
rescaled by a factor 0.88 to account for mass uncertainties and missing entropy.
The phonon contribution estimated from a high-$T$ fit (purple dashed line)
is negligible below 10~K.}
\label{Fig-chi_cv}
\end{figure}

The DC magnetic susceptibility measured in a SQUID in a magnetic field of 5~T
exhibits a Curie-Weiss behavior $\chi_{\rm DC}(T)=C/(T-\theta_{CW})$
above $\sim$50~K with a weak ferromagnetic effective coupling, $\theta_{CW}=9.5 \pm 1$~K
[Fig.~\ref{Fig-chi_cv}(a)].
For $T \lesssim 50$~K, the susceptibility becomes Curie-like,
with neither any sign of a magnetic transition nor a FC/ZFC opening.
A similar Curie-tail is observed in many other compounds
and is often attributed to a defect contribution
which exceeds the intrinsic susceptibility~\cite{Olariu08, Quilliam11}.
The use of a local probe such as NMR is decisive in this context:
We find that the local susceptibility \Xloc,
extracted from the $^{35}$Cl NMR shift
of an oriented powder sample
with magnetic fields up to 13~T applied perpendicular to the kagom\'e planes,
scales with \XDC\ over the whole temperature range 2--287 K.
This clearly shows that isolated spin defects are not present in
kapellasite in contrast with the behavior seen in
herbertsmithite, where Cu/Zn mixing creates paramagnetic-like spins.
Thus, the intrinsic susceptibility of kapellasite displays genuine
divergent behavior pointing to a gapless spin state, rather than the
downturn seen in other KAFM compounds.

% FIG. 3. Thermo

A reasonable first theoretical description of kapellasite,
which is deep in the Mott phase with well-defined localized spins,
is the Heisenberg Hamiltonian
\begin{equation}
\mathcal{H} = \sum_{{\langle i,j\rangle}_\alpha} J_\alpha \, {\mathbf S}_i \cdot {\mathbf S}_j ,
\label{EqHberg}
\end{equation}
where the exchange integrals $J_\alpha$
are defined in Fig.\ \ref{FigLattice}(a).
We use $n$th order high-temperature (HT) series expansions
\cite{Bernu01,Misguich05}
to fit bulk \XDC\ data
using $J_1$--$J_2$ or  $J_1$--$J_d$ models,
following suggestions that $J_d$ is significant in kapellasite \cite{Janson08}.
In both cases, $J_1$ comes out ferromagnetic
in competition with  further-neighbor antiferromagnetic interactions
[see red points in Fig.\ \ref{FigLattice}(b)],
which explains the low value of $\theta_{CW}$.
Since a $n$th order HT series expansion depends on combinations of up to $n$th points correlation functions that describe the short-range arrangements of the spins,
it is not surprising that the two fits fall in the same phase.
The fits are very stable, and moving the parameters from the cuboc2 phase towards the ferromagnetic one, say,
leads rapidly to a poor agreement between theory and experiment.
The fit of the $J_1$--$J_d$ model ($n$=11) is qualitatively better
than the $J_1$--$J_2$ model
and gives a nearest-neighbor interaction $J_1=-15.0(4)$ K
and an ``across-hexagon'' interaction $J_d=12.7(3)$ K.
The ferromagnetic nearest-neighbor exchange ($J_1$)  highlights the difference between kapellasite and its polymorph herbertsmithite, and can be traced back to
the Cu--$\mu_3$OH--Cu bonding angle  being $\sim$13$^\circ$ smaller in kapellasite \cite{Colman10,Wills07}.
The low-$T$ peak in the specific heat is also well captured by a HT-series expansion using the above $J_1$--$J_d$ parameters, as shown in Fig.~\ref{Fig-chi_cv}(b).

Inelastic neutron scattering (INS) measurements were performed at $0.1<T<300$~K
on the neutron time-of-flight spectrometer IN5 at the ILL,
using incident neutron energies of 0.82, 3.27, and 20.45 meV.
A 93\% deuterated powder sample of mass 3~g was held in an annular aluminium can
and thermalized by helium exchange gas.
Standard data reduction \cite{lamp} gave the neutron scattering function
 \SQE, shown in Fig.~\ref{FigMap}.
The magnetic scattering,
which dominates the nuclear (phonon) scattering
for the energies and wave vectors studied in this work,
is quasi-elastic and essentially without dispersion.

% FIG. 4. Neutrons
\begin{figure}
\includegraphics[width=0.98\columnwidth]{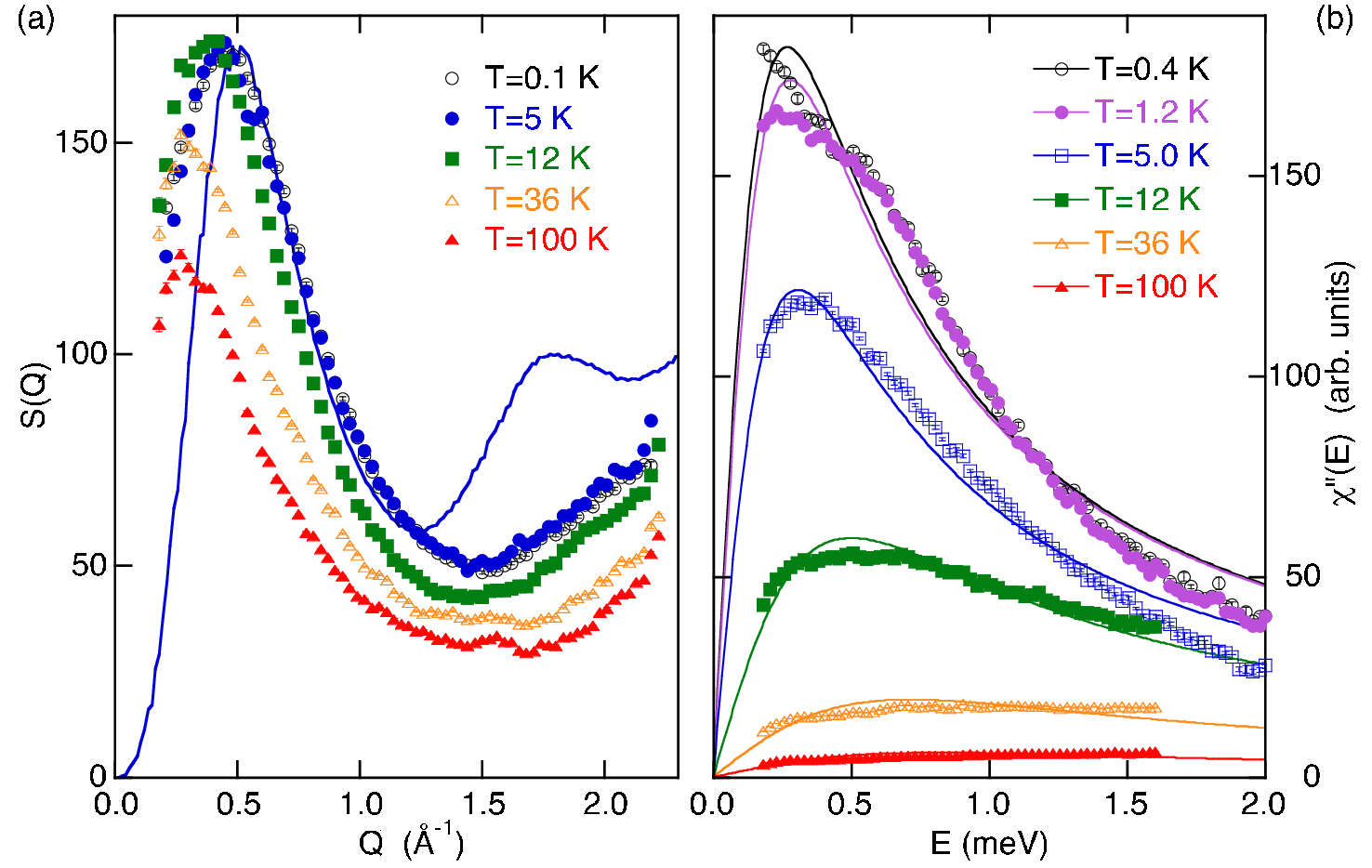}
\caption{(Color online)
(a) Wave-vector dependence of the inelastic magnetic scattering, \SMQ,
integrated over the energy range $0.4\!<\!E\!<\!0.8$ meV.
Symbols are experimental data for different temperatures
(statistical errors are smaller than the symbol size)
and the line is the theoretical SBMF calculation at $T=5$ K.
(b) Energy dependence of the imaginary part
of the magnetic dynamic susceptibility, \XE,
integrated over the $Q$ range $0.4\!<\!Q\!<\!0.8$ \Ang\
for different temperatures. The energy resolution is 0.11 meV.
The data show deviations from a quasi-elastic Lorentzian (lines) at low temperatures.}
\label{FigSQE}
\end{figure}

The inelastic magnetic structure factor, \SMQ,
obtained by integrating \SQE\ over the energy range $0.4<E<0.8$ meV
and correcting for the magnetic form factor of the Cu$^{2+}$ ions,
is shown in Fig.\ \ref{FigSQE}(a).
The measured \SMQ\ has initially only a weak temperature dependence,
but appears to shift to slightly lower $Q$ values for $T>15$ K.
At even higher temperatures the structure gradually disappears,
but short-range correlations persist up to at least 100 K.
At low temperatures, the broad peak near $Q$=0.5 \Ang\ [see Fig.\ \ref{FigSQE}(a)]
is centered at a much smaller value than in other kagom\'e systems.
The observed peak position corresponds neither to the
$\Gamma$ point (${\bf k}=0$) in the Brillouin zone
nor to the K point {\bf k}=(1/3,1/3) of the $\sqrt 3\times\sqrt 3$ structure,
but is close to the M point, {\bf k}=(1/2,0).
The M point on the kagom\'e lattice corresponds to a {\it unique} spin arrangement
among all possible regular spin states \cite{Messio11} of the Hamiltonian (\ref{EqHberg}), called cuboc2 (see inset of Fig.\ \ref{FigMap} and Ref.~\cite{movie}),
and is fully consistent with the analysis of the HT spin susceptibility and the classical phase diagram of the derived Hamiltonian [see Fig.\ \ref{FigLattice}(b)].
These two independent measurements, in the high- and low-temperature regimes,
give strong support for that the \textit{short-range} correlations of kapellasite are of non-coplanar cuboc2 type.

To confirm the relevance of the cuboc2 state for quantum spins,
we have explored the impact of quantum fluctuations on the classical system
by employing  a Schwinger-boson mean-field
approach (SBMF) \cite{Messio10}
of the Hamiltonian (\ref{EqHberg}),
using the values of $J_1$ and $J_d$ extracted from \XDC\
and a spin value of $S=0.5$.
The powder-averaged structure factor $S(Q)$
calculated at $T$=5 K
for the same energy range as in the experiment
is displayed as a line in Fig.\ \ref{FigSQE}(a),
with only the overall scale factor adjusted.
The position of the first peak in $S(Q)$ at $Q=0.5$~\Ang\ is in good agreement with experiment,
as is the weak intensity near $Q=1$~\Ang.
These two features are the main signatures of the cuboc2 phase.
Classical simulations show that replacing Cu by Zn on 27\% of the kagom\'e sites
broaden the structure factor so that the peaks at large $Q$s are partly blurred out
while the first peak in the powder-averaged $S(Q)$ near $Q=0.5$~\Ang\ appears at slightly smaller $Q$s \cite{Classim}.
Adding quantum fluctuations to this dilution effect could lead to even better agreement with INS data.

The spin dynamics studied by INS
displays a smooth continuum of excitations up to at least 2 meV
with no discernible gap and no sign of damped spin waves,
see Fig.\ \ref{FigMap}.
We extract the energy dependence by integrating \SQE\
over the wave vector range $0.4<Q<0.8$ \Ang,
where the magnetic scattering is strong and the nuclear scattering negligible,
and divide by the temperature factor $1-\exp(-E/k_BT)$
to obtain the imaginary part of the magnetic dynamic susceptibility, \XE,
shown in  Fig.~\ref{FigSQE}(b).
At high temperatures,
\XE\ is well described by a quasi-elastic Lorentzian,
$\XE= \chi' E  \Gamma/(E^2 + \Gamma^2)$,
similar to the classical ($S\!=\!5/2$) KAFM deuteronium jarosite \cite{Fak08}.
For $T > 5$ K,
the line width (inverse relaxation rate) increases as $\Gamma(T)\propto T^{1/3}$ with temperature
while the staggered static susceptibility decreases as $\chi'(T)\propto T^{-2/3}$.
At lower temperatures, below 5 K,
the shape of \XE\ deviates from a single Lorentzian,
which signals the onset of quantum spin-liquid correlations.
However, kapellasite does not show the $E^{-\alpha}$ behavior
of pyrochlore slabs \cite{Broholm90} nor the $E/T$ scaling of herbertsmithite~\cite{Helton07,Helton10}.
This gives further support for kapellasite not being close to a quantum critical point.
A full understanding of the spin liquid dynamics of kapellasite is not yet at hand and will be discussed further below.

% FIG. 5. uSR
\begin{figure}%[h!]
\includegraphics[width=\columnwidth,trim= 0.75cm 0.75cm 0.75cm 0.75cm,keepaspectratio]{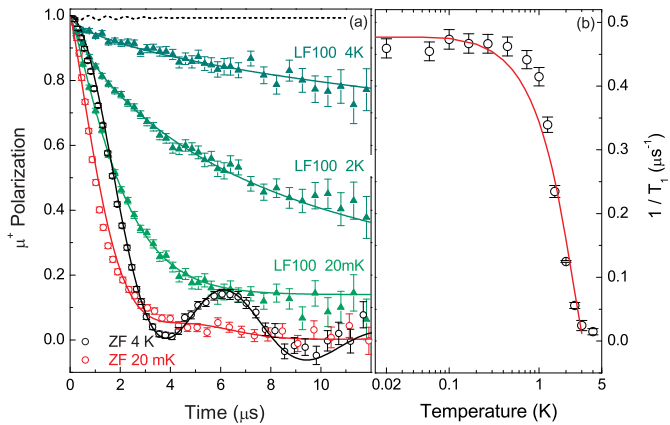}
\caption{(Color online) (a) Time dependent muon polarization
in zero field (open symbols)
and longitudinal field (closed symbols)
demonstrate the absence of spin freezing.
Solid lines are fits to Eq.~(\ref{EqPol})
while the dashed line represents the expected LF polarization at 100~G
for frozen spins at 20~mK.
(b) Temperature dependence of the relaxation rate $1/T_1$
from $\mu$SR measurements (circles)
and SBMF calculations (red line).}
\label{FiguSR}
\end{figure}

Further insight into the low-$T$ state of kapellasite
is provided by our $\mu$SR experiments,
which were performed on a non-oriented powder sample
in zero field (ZF) and longitudinal applied field (LF)
at the ISIS and PSI facilities down to 20~mK.
The evolution of the muon polarization is plotted in Fig.\ \ref{FiguSR}(a),
both in zero field and under $H\!=\!100$~G from 4~K to 20~mK.
The relaxation of the muon polarization can be fitted in the whole $T$ range
to the sum of the local responses of two muon sites,
known to be located nearby OH$^{-}$ and Cl$^{-}$~\cite{Mendels07},
\begin{equation}
    P(t) = f P_\mathrm{OH}(t)e^{-\lambda_\mathrm{OH} t}
    + (1-f)K_\mathrm{Cl}(t)e^{-\lambda_\mathrm{Cl} t} ,
\label{EqPol}
\end{equation}
where $f \sim 0.86$ is the $T$-independent fraction
of muons on the OH$^{-}$ site
and $\lambda_\mathrm{OH/Cl}$ is the dynamical electronic relaxation rate
probed at the OH$^{-}$/Cl$^{-}$ site.
Both $P_\mathrm{OH}(t)$ and the Kubo-Toyabe function $K_\mathrm{Cl}(t)$
are related to the nuclear static fields \cite{Mendels07}
which are $T$ independent and were fixed to their $T=4$~K values.

Any spin freezing can be ruled out based on the following arguments.
First, in a fully static scenario,
the low-$T$ relaxation should be ascribed to a distribution of static internal fields
with a width $\Delta H\sim 6$~G
which should be completely decoupled
in an applied LF of 100~G.
The dashed line in Fig.\ \ref{FiguSR}(a) shows that this is clearly not the case.
Second, the ZF polarization at 20 mK relaxes to zero,
not showing the long-time $1/3$-component characteristic of frozen moments.
Finally, the $T$ dependence of the ZF polarization is well described by Eq.~(\ref{EqPol})
in the whole temperature range from 20~K to 20~mK
with only one adjustable parameter, $\lambda_{OH}$,
to account for the Cu spin dynamics.
No electronic relaxation ($e^{-\lambda_\mathrm{Cl} t} \sim 1$) was found at the Cl site,
which is further away from the magnetic planes.

To study the slow spin dynamics,
an applied  field of 100~G along the $\mu^{+}$ polarization
decouples the static nuclear magnetism so that Eq.~(\ref{EqPol}) becomes
$P(t) = fe^{-\lambda t} + (1-f )$,
and the spin-lattice relaxation rate $1/T_1= \lambda_{OH}$ is probed directly
[Fig.\ \ref{FiguSR}(b)].
Above 4~K, the $S\!=\!1/2$ kagom\'e spins are in a rapidly fluctuating regime.
From 3~K to 0.8~K we observe a slowing down of spin fluctuations,
which occurs in the same $T$-range as the change in dynamics in neutron experiments
and coincides with the low-$T$ specific heat peak.
This indicates an enhancement of the short-range correlations at low $T$,
and the corresponding energy scale is quite low as a
result from a competition between the exchange interactions.
Finally, persistent fluctuations are observed down to 20~mK.
This behavior points to a gapless spin-liquid state with an upper bound
of a hypothetical spin gap of 20~mK, i.e.,\ $10^{-3}J_1$.
Within our SBMF approach, the \uSR\  spin-lattice relaxation rate can be calculated as
$1/T_1\propto T [\chi''(\omega_L, T)/\omega_L]$,
where $\omega_L$ is the muon Larmor frequency.
The theoretical results, also shown in Fig.\ \ref{FiguSR}(b),
are in excellent qualitative agreement with the experiment,
demonstrating that the local dynamics of kapellasite
is well captured by the SBMF  calculations at low frequencies.
We note that the SBMF calculations for both the muon relaxation rate and \SQE\
give essentially the same predictions for both the $J_1$--$J_d$ and $J_1$--$J_2$ models.
This suggests that the importance lies not in the exact values of the exchange parameters used,
but rather in which part of the phase diagram kapellasite is located, and this is unambiguously determined by the HT expansion.

At higher frequencies, our SBMF calculations predict well-defined spin waves for $T<3$ K
with energies around 0.5 meV.
We expect that dilution of the magnetic sites reduces the spin-spin correlation length and broadens the modes. However, \SQE\  would still display reminiscences of spin waves, resulting in a non-monotonous energy response for certain $Q$ values (such as the magnetic zone boundary), which is clearly not the case in kapellasite, where the energy dependence of the intensity at all $Q$s decrease monotonously.  It thus appears clear that our a priori choice of bosonic excitations is too restrictive to capture both the continuum of spinon excitations and the absence of a gap.
This  indeed calls for further theoretical explorations.  A fermionic chiral approach would plausibly lead both to gapless and diffuse excitations and to a better description of the inelastic neutron spectrum.  Zero-temperature fermionic approaches of the pure antiferromagnetic Heisenberg Hamiltonian \cite{Ran07,Iqbal11a,Iqbal11b}  have been extremely interesting and similar work should be extended to the present model with competing interactions,
with the aim to describe both chiral physics and thermodynamic properties.
Whether the dilution of Cu ions is strong enough to kill the predicted low-$T$ chiral phase  expected  with this local spin arrangement \cite{Domenge08}
is also an interesting question.

This work was supported in part by the
European Commission under the 6th Framework Programme Contract
No.\ RII3-CT-2003-505925, ANR Grant No.\ ANR-09-JCJC-0093-01 and ARCUS Ile de France-Inde.
We thank C. Baines for assistance at PSI,
D. Dragoe for the ICP analysis,
and P. Viot and F. Brieuc for their collaboration on the simulation of disorder.


\begin{thebibliography}{11}

\bibitem{Balents10} L. Balents,
%{\it Spin liquids in frustrated magnets},
Nature {\bf 464}, 199 (2010).

\bibitem{Shores05} M. P. Shores, E. A. Nytko, B. M. Bartlett, and D. G. Nocera,
%{\it A structurally perfect $S=1/2$ kagom\'e antiferromagnet},
J. Am. Chem. Soc. {\bf 127}, 13462 (2005).

\bibitem{Mendels07} P. Mendels \etal,
%F. Bert, M. A. de Vries, A. Olariu, A. Harrison, F. Duc, J. C. Trombe, J. S. Lord, A. Amato, and C. Baines,
%{\it Quantum Magnetism in the Paratacamite Family: Towards an Ideal Kagom\'e Lattice }
Phys. Rev. Lett. \textbf{98}, 077204 (2007).

\bibitem{Helton07} J. S. Helton \etal,
%K. Matan, M. P. Shores, E. A. Nytko, B. M. Bartlett, Y. Yoshida, Y. Takano, A. Suslov, Y. Qiu, J.-H. Chung, D. G. Nocera, and Y. S. Lee,
%{\it Spin dynamics of the spin-1/2 kagome lattice antiferromagnet ZnCu$_3$(OH)$_6$Cl$_2$},
Phys. Rev. Lett. {\bf 98},  107204  (2007).

\bibitem{Olariu08} A. Olariu \etal,
%P. Mendels, F. Bert, F. Duc, J. C. Trombe, M. de Vries, and A. Harrison,
%{\it $^{17}$O NMR Study of the intrinsic magnetic susceptibility and spin dynamics of the quantum kagome antiferromagnet ZnCu$_3$(OH)$_6$Cl$_2$},
Phys. Rev. Lett. {\bf 100}, 087202 (2008).

\bibitem{deVries09} M. A. de Vries \etal,
%J. R. Stewart, P. P. Deen, J. O. Piatek, G. J. Nilsen, H. M. R\o nnow, and A. Harrison,
%{\it Scale-free antiferromagnetic fluctuations in the $S=1/2$ kagome antiferromagnet Herbertsmithite},
Phys. Rev. Lett. {\bf 103}, 237201 (2009).

\bibitem{Mendels10} P. Mendels and F. Bert,
%{\it Quantum kagome antiferromagnet ZnCu$_3$(OH)$_6$Cl$_2$},
J. Phys. Soc. Jpn. {\bf 79},  011001  (2010).

\bibitem{Jeong11} M. Jeong \etal,
%F. Bert, P. Mendels, F. Duc, J.-C. Trombe, M.A. de Vries, and A. Harrison
%{\it Field-induced freezing of a quantum spin liquid on the kagome lattice},
Phys. Rev. Lett. {\bf 107}, 237201 (2011).

\bibitem{Ran07}
Y. Ran,  M. Hermele, P. A. Lee, and X. G. Wen,
%{\it Projected-wave-function study of the spin-1/2 Heisenberg model on the kagom\'e lattice},
Phys. Rev. Lett. {\bf 98}, 117205 (2007).

\bibitem{Yan11} S. Yan, D. A. Huse, and S. R. White,
%{\it Spin-Liquid Ground State of the S = 1/2 Kagome Heisenberg Antiferromagnet},
Science {\bf 332}, 1173 (2011).

\bibitem{Lu11}
Y.-M. Lu, Y. Ran, and P. A. Lee,
%{\it $Z_2$ spin liquids in the $S$=1/2 Heisenberg model on the kagome lattice: A projective symmetry-group study of Schwinger fermion mean-Þeld states},
Phys. Rev. B {\bf 83}, 224413 (2011).

\bibitem{Messio12}
L. Messio, B. Bernu, and C. Lhuillier,
%{\it Kagome antiferromagnet: A chiral topological spin liquid?},
Phys. Rev. Lett. {\bf 108}, 207204 (2012). % arXiv:1110.5440v2.

\bibitem{Colman08} R. H. Colman, C. Ritter, and A. S. Wills,
%{\it Toward perfection: Kapellasite, Cu$_3$Zn(OH)$_6$Cl$_2$, a new model $S=1/2$ kagome antiferromagnet},
Chem. Mater. {\bf 20}, 6897 (2008).

\bibitem{Colman10} R. H. Colman, A. Sinclair, and A. S. Wills,
%{\it Comparison between Haydeeite, $\alpha$-Cu$_3$Mg(OD)$_6$Cl$_2$, and Kapellasite, $\alpha$-Cu$_3$Zn(OD)$_6$Cl$_2$, isostructural $S$=1/2 kagome magnets},
Chem. Mater. {\bf 22}, 5774 (2010).

\bibitem{Domenge05} J.-C. Domenge,
P. Sindzingre, C. Lhuillier, and  L. Pierre,
%{\it Twelve sublattice ordered phase in the $J_1$-$J_2$ model on the kagome lattice},
Phys. Rev. B {\bf 72}, 024433 (2005).

\bibitem{Kermarrec12} E. Kermarrec \etal,
{\it in preparation.}

\bibitem{Quilliam11} J. A. Quilliam, %\etal,
F. Bert, R. H. Colman, D. Boldrin, A.~S.~Wills, and P. Mendels,
%{\it Ground state and intrinsic susceptibility of the kagome antiferromagnet vesignieite as seen by $^{51}$V NMR},
Phys. Rev. B {\bf 84}, 180401(R) (2011).

\bibitem{Bernu01}  B. Bernu and G. Misguich,
%{\it Specific heat and high-temperature series of lattice models: Interpolation scheme and examples on quantum spin systems in one and two dimensions},
Phys. Rev. B {\bf 63},  134409 (2001).

\bibitem{Misguich05} G. Misguich and B. Bernu,
%{\it Specific heat of the S=1/2 Heisenberg model on the kagome lattice: High-temperature series expansion analysis},
Phys. Rev. B {\bf 71},  014417 (2005).

\bibitem{Janson08} O. Janson, J. Richter, and H. Rosner,
%{\it Modified kagome physics in the natural spin=1/2 kagome lattice systems: Kapellasite Cu$_3$Zn(OH)$_6$Cl$_2$ and Haydeeite Cu$_3$Mg(OH)$_6$Cl$_2$},
Phys. Rev. Lett. {\bf 101}, 106403 (2008).

\bibitem{Wills07} A. S. Wills and J.-Y. Henry,
%{\it On the crystal and magnetic ordering structures of clinoatacamite, $\gamma$-Cu$_2$(OD)$_3$Cl, a proposed valence bond solid},
J. Phys.: Condens. Matter {\bf 20}, 472206 (2007).

\bibitem{lamp}  {\sf LAMP},
%the large array manipulation program,
http://www.ill.fr/data\_treat/lamp/lamp.html.

\bibitem{Messio11} L. Messio, C. Lhuillier, and G. Misguich,
%{\it  Lattice symmetries and regular magnetic orders in classical frustrated antiferromagnets},
Phys. Rev. B {\bf 83},  184401 (2011).

\bibitem{movie} See Supplemental Material at http://link.aps.org/supplemental/10.1103/PhysRevLett.109.037208 for a three-dimensional animation of the cuboc2 structure.

\bibitem{Messio10} L. Messio, O. C\'epas, and C. Lhuillier,
%{\it Schwinger-boson approach to the kagome antiferromagnet with Dzyaloshinskii-Moriya  interactions: Phase diagram and dynamical structure factors},
Phys. Rev. B {\bf 81}, 064428 (2010).

\bibitem{Classim} See Supplemental Material at [URL2] for details on the classical simulation.

\bibitem{Fak08} B. F{\aa}k, % \etal,
F. C. Coomer, A. Harrison, D. Visser, and M. E. Zhitomirsky,
%{\it Spin-liquid behavior in a kagom\'e antiferromagnet: Deuteronium jarosite},
Europhys. Lett.  {\bf 81}, 17006 (2008).

\bibitem{Broholm90} C. Broholm,
G. Aeppli, G. P. Espinosa, and A. S. Cooper,
%{\it Antiferromagnetic fluctuations and short-range order in a kagom\'e lattice},
Phys. Rev. Lett. {\bf 65}, 3173 (1990).

\bibitem{Helton10} J. S. Helton \etal,
%K. Matan, M. P. Shores, E. A. Nytko, B. M. Bartlett, Y. Qiu, D. G. Nocera, and Y. S. Lee,
%{\it Dynamic Scaling in the Susceptibility of the Spin-$1/2$ Kagome Lattice Antiferromagnet Herbertsmithite},
Phys. Rev. Lett. {\bf 104}, 147201 (2010).

\bibitem{Iqbal11a} Y. Iqbal, F. Becca, and D. Poilblanc,
%{\it Valence-bond crystal in the extended kagome spin-1/2 quantum Heisenberg antiferromagnet: A variational Monte Carlo approach},
Phys. Rev. B {\bf 83}, 100404(R) (2011).

\bibitem{Iqbal11b} Y. Iqbal, F. Becca, and D. Poilblanc,
%{\it Projected wave function study of $Z_2$ spin liquids on the kagome lattice for the spin-1/2 quantum Heisenberg antiferromagnet},
Phys. Rev. B {\bf 84}, 020407(R) (2011).

\bibitem{Domenge08} J.-C. Domenge, % \etal,
C. Lhuillier, L. Messio, L. Pierre, and P.~Viot,
%{\it Chirality and $Z_2$ vortices in a Heisenberg spin model on the kagome lattice},
Phys. Rev. B {\bf 77}, 172413 (2008).


\end{thebibliography}
\end{document}